\documentstyle[prd,aps,twocolumn,epsf,floats,amsfonts,amssymb,amsmath]{revtex}
\newcommand{\bea}{\begin{eqnarray}}\newcommand{\eea}{\end{eqnarray}}
\newcommand{\ba}{\begin{array}}\newcommand{\ea}{\end{array}}
\newcommand{\bit}{\begin{itemize}}\newcommand{\eit}{\end{itemize}}
\newcommand{\ben}{\begin{enumerate}}\newcommand{\een}{\end{enumerate}}
\newcommand{\bib}{\bibitem}
\newcommand{\lab}{\label}
\newcommand{\lan}{\langle}
\newcommand{\lf}{\left}

\newcommand{\pa}{\partial}\newcommand{\ran}{\rangle}

\newcommand{\ri}{\right}

\newcommand{\ga}{\gamma}\newcommand{\Ga}{\Gamma}
\newcommand{\De}{\Delta}
\newcommand{\ep}{\epsilon}

\newcommand{\ka}{\kappa}

\newcommand{\Om}{\Omega}\newcommand{\I}{{_I}}
\newcommand{\II}{{_{I\!I}}}

\begin{document}
\twocolumn[\hsize\textwidth\columnwidth\hsize\csname@twocolumnfalse\endcsname

\title{Dissipation and Quantization}

\author{Massimo Blasone${}^{\sharp \flat }$, Petr Jizba${}^{\natural}$,
and Giuseppe Vitiello${}^{\flat }$ \vspace{3mm}}

\address{${}^{\sharp}$   Blackett  Laboratory, Imperial  College, London
SW7 2BZ, U.K.
\\ [2mm] ${}^{\natural}$FNSPE, Czech Technical
University, Brehova 7, 115 19 Prague 1, Czech Republic
\\ [2mm] ${}^{\flat }$
Dipartimento di  Fisica and INFN, Universit\`a di Salerno, 84100
Salerno, Italy
\vspace{2mm}}


\maketitle

\begin{abstract}
We show that
the dissipation term in the Hamiltonian for a couple of classical
damped-amplified oscillators manifests itself as a geometric
phase and is actually responsible for the appearance
of the zero point
energy in the quantum spectrum of the 1D linear harmonic oscillator.
We also discuss the thermodynamical
features of the system. Our work has been inspired by 't Hooft proposal
according to which
information loss in certain classical systems may lead to
``an apparent quantization of the orbits which
resembles the quantum structure seen in the real world".

\end{abstract}

\vspace{8mm}]



Recently \cite{'tHooft:1999gk,erice,'tHooft:1988sx} Gerard 't Hooft has
discussed classical, deterministic, dissipative models and has shown
that constraints imposed on the solutions which introduce information
loss provide bounded Hamiltonians and lead to {\em ``an apparent
quantization of the orbits which resembles the quantum structure seen
in the real world''}\cite{erice}. 't Hooft's conjecture is that the
dissipation of information which would occur at Planck scale in a
regime of completely deterministic dynamics would play a r\^ole in the
quantum mechanical nature of our world.

The purpose of this Letter is
to present an example of dissipation in a classical system  which
explicitly leads, under suitable conditions, to a quantum behavior. We
show that the dissipation term in the Hamiltonian for a couple of
classical damped-amplified oscillators manifests itself as a geometric
phase and is actually responsible for the appearance of the zero point
energy in the quantum spectrum of the 1D linear harmonic oscillator. We
also discuss the thermodynamical features of the system.

The considered system of
two oscillators, one damped, the other one amplified, has revealed
to be an useful playground for the study of several problems of physical
interest, such as the study of phase coherence in quantum Brownian motion
\cite{Blasone:1998xt}, the study of topologically massive gauge theories in
the infrared region in $2+1$ dimensions \cite{Blasone:1996yh}, the
Chern-Simons-like dynamics of Bloch electrons in solids
\cite{Blasone:1996yh}, and exhibits features also common to
the structure of two-dimensional gravity models \cite{Cangemi}.

By imposing a condition of ``adiabaticity''
on such a classical system of two dissipative oscillators,
we obtain a one--dimensional quantum oscillator with
zero point energy  originating in the geometric phase
of the  classical system. Thus our conclusions seem to support
't Hooft's analysis.

Let us first briefly outline 't~Hooft's scenario.
He considered  Hamiltonians of
the form $H = \sum_{i}p_{i}\, f_{i}(q)$, where $f_{i}(q)$ are
non--singular functions of the canonical coordinates $q_{i}$.
The crucial point is that equations
for the $q$'s (i.e. $\dot{q_{i}} = \{q_{i}, H\} =
f_{i}(q)$) are decoupled from the conjugate momenta
$p_i$. This means \cite{erice} that the system
can be described deterministically even when expressed in terms of
operators acting on some functional space of states, such as the Hilbert
space. This is possible because it exists, for such systems, a complete
set of observables which Poisson commute at all times. They are called
{\em beables} \cite{Bell:1987hh}. Of course, such a description in terms
of operators and Hilbert space, does not implies {\em per se}
quantization of the system.
Quantization is in fact achieved only as a consequence of the
dissipation of information, according to 't Hooft's scenario.

The above mentioned class of Hamiltonians is, however,
not bounded from below.
This might be cured by splitting $H$ as\cite{erice}:  $H = H_1 -
H_2$, with
\bea
&&H_1 = \frac{1}{4\rho}\left( \rho +
H\right)^{2}\;\; ,\;\;\;
H_2 = \frac{1}{4\rho}\left( \rho -
H\right)^{2}\, ,  \label{1.5}
\eea
and  $\rho$ a certain time--independent, positive function of $q_{i}$.
As a result, $H_1$ and $H_2$ are positively
(semi)definite and $ \{H_1, H_2\} =
\{\rho , H \} =  0\,$ .

The non--linear nature of the evolution could lead to
integral curves (presumably even chaotic \cite{Biro:1999bh})
in the configuration space with
one--dimensional attractive trajectories (e.g. limit cycles).
States that are initially different
may evolve into the same final state: attractive trajectories may then
be treated as  distinct equivalence classes.
`t Hooft has observed\cite{'tHooft:1988sx}
that the dynamics of classical deterministic systems
may be mathematically formulated in terms of the unitary evolution
operator acting on the Hilbert space of states $|\psi\ran$
and expressed in the form of
a Schr\"odinger--like equation.  Once again, we stress that
this is yet not
equivalent to quantization, but it is a pure
mathematical construct allowing one
to get a statistical inference about the system in question.

To get the lower bound for the Hamiltonian one thus imposes
the constraint condition onto the Hilbert space:
\bea
H_2|\psi \ran = 0\, ,
\label{3.8}
\eea
which projects out the states responsible for the
negative part of the spectrum. In the deterministic language
this means that one gets rid of the unstable trajectories.

What we present here is an explicit realization of 't Hooft mechanism,
with the further step of discovering a connection between the zero
point energy and the geometric phase. The model we consider is,
however, different from 't Hooft model. Nevertheless, the Hamiltonian
of our model belongs to the same class of the Hamiltonians considered
by 't Hooft, as we explicitly show below.

We start our discussion by considering a system of
1D damped and amplified harmonic oscillators:
\bea \lab{eqx}
m \ddot x + \ga \dot x + \ka x = 0  \, ,
\\ \lab{eqy}
m \ddot y - \ga \dot y + \ka y = 0 \, ,
\eea
respectively. The $y$-oscillator is the time--reversed
image of the $x$-oscillator. The corresponding Hamiltonian  reads
\bea \lab{xyham}
H=\frac{1}{m} p_{x} p_{y} + \frac{1}{2m}\ga  \lf( y
 p_{y} - x p_{x} \ri) + \lf( \ka -  \frac{\ga^{2}}{4 m}\ri) x y \, ,
\eea
with $\, p_{x}  = m \dot y - \frac{1}{2} \ga y$ ;  $p_{y} = m \dot x +
\frac{1}{2} \ga x \,$.

In order to show that $H$ of Eq.(\ref{xyham}) belongs to the
class of Hamiltonians above mentioned, we introduce the following
notation:
$x_1 = (x + y)/\sqrt 2$, $x_2 = (x - y)/\sqrt 2$
and  $\, p_{1}  = m {\dot x}_{1} + \frac{1}{2} \ga {x_2}$ ;  $p_{2} =
- m {\dot x}_{2} -
\frac{1}{2} \ga {x_1} \,$.
We also put \cite{Blasone:1996yh} $ x_1 =  r \cosh u $,
$ x_2 =  r \sinh u$, and
\bea\lab{ca} {\cal C} &=& \frac{1}{4 \Om m}\lf[ \lf(p_1^2  - p_2^2\ri)+
m^2\Om^2 \lf(x_1^2 -  x_2^2\ri)\ri] \\  \lab{cc} &=&\frac{1}{4 \Om m}\lf[
p_r^2  - \frac{1}{r^2}p_u^2 + m^2\Om^2 r^2\ri]\, ,
\\ \lab{j2a}
J_2&=& \frac{m}{2}\lf[\lf( {\dot x}_1 x_2 - {\dot x}_2
x_1 \ri) - {\Ga} r^2 \ri]
= \frac{1}{2}\, p_u\, ,
\eea
where $\Ga = \ga/ 2 m$, $\Om = \sqrt{\frac{1}{m}
(\ka-\frac{\ga^2}{4m})}$, with $\ka >\frac{\ga^2}{4m} $.

Eq.(\ref{xyham}) can be then rewritten as:
\bea \lab{pqham}
H &=& \sum_{i=1}^2p_{i}\, f_{i}(q)\,,
\eea
with $f_1(q)=2\Om$,  $f_2(q)=-2\Ga$, provided we use the canonical
transformation \cite{goldstein}:
\begin{eqnarray}
&&q_{1} = \int
\frac{dz \;\; m\Omega}{\sqrt{4 J_{2}^{2} + 4m\Omega {\cal{C}
} z - m^{2}\Omega^{2} z^{2}}}\, ,\\
&&q_{2} = 2u + \int \frac{dz}{z} \,
\frac{2J_{2}}{\sqrt{4 J_{2}^{2} + 4m\Omega {\cal{C}}
z - m^{2}\Omega^{2} z^{2}}}\, ,\\
&&p_1 = {\cal C}\, , \;\;\;\;\
p_2 = J_2 \, ,
\label{can1}
\end{eqnarray}
with $z=r^2$. One has $\{q_{i},p_i\} =1$, and the other Poisson
brackets vanishing.
The above
notation reminds one that the structure of the Hamiltonian is the one of
$su(1,1)$ \cite{Celeghini:1992yv}
(${\cal C}$ is the Casimir operator and $J_2$
one  of its  generators).

Thus from the very definition  $J_2$ and
${\cal C}$  are beables. Yet also $q_1$ and $q_2$ are beables
(although members of a different Lie sub-algebra than their
conjugates)  as it can be directly seen from the
Hamiltonian (\ref{pqham}). The fact that a given dynamics admits more
than one Lie-subgroup of beables is by no means generic and here it is
only due to a special structure of $H$.  In such cases it is quite
intriguing question which set of beables should be used.
We intend here to exploit the outcomes of the
$J_2$, $C$ set.

We put  $H = H_{\I} - H_{\II}$, with
\bea
&&H_{\I} = \frac{1}{2 \Om {\cal C}} (2 \Om {\cal C}
- \Ga J_2)^2\;\;\; ,\;\;\;\;
H_{\II} = \frac{\Ga^2}{2 \Om {\cal C}} J_2^2\, .  \label{split}
\eea
Of course, only  nonzero $r^{2}$
should be taken into account in order for $\cal C$ to be invertible.
Note that ${\cal C}$ is a constant of motion (being the Casimir
operator): this
ensures that once it has been chosen to be positive, as we do from
now on, it will remain such at all times.

We  now implement the constraint
\bea
\label{thermalcondition}\quad J_2
|\psi\ran = 0\, ,
\eea
which defines the physical states.
Although the system (\ref{pqham}), i.e.(\ref{xyham}), is deterministic,
$|\psi\ran$
is not an eigenvector of $u$ ($u$ does not commute with $p_u)$.
Of course,
if one does not use the operatorial formalism to describe
our system, then $p_u=0$ implies $u = -\frac{\ga}{2 m} t$. This point
illustrate the specific feature of the description of a classical
deterministic system in terms of beables, i.e. of operators commuting at
all times: $u$ is not a beable.

It should be also mentioned that the  constraint
(\ref{thermalcondition}) automatically implies that the corresponding
conjugate variable $q_2$ is unbounded in the physical states
$|\psi\rangle$ ( see Eq.(11) ).  It should be born
in mind that such a behavior of conjugate variables reflects into a
typical feature of quantum theory as it assures a compatibility with
the Heisenberg uncertainty relation. Eq.(\ref{thermalcondition})
implies
\bea \lab{17}
H |\psi\ran= H_\I |\psi\ran=  2\Om {\cal C}|\psi\ran
= \left( \frac{1}{2m}p_{r}^{2} + \frac{K}{2}r^{2}\right) |\psi \ran \, ,
\eea
where  $K\equiv m \Om^2$. $H_\I$ thus reduces to the Hamiltonian for the
linear harmonic oscillator $\ddot{r} + \Om^2 r =0 $.
The physical states are even with respect to
time-reversal ($|\psi(t)\ran  = |\psi(-t)\ran$) and
periodical with period $\tau = \frac{2\pi}{\Omega}$.

Let us  write the generic state $|\psi(t)\ran_{H}$ as
\bea\lab{eqt0}
|\psi(t)\ran_{H} = {\hat{T}}\lf[ \exp\lf( \frac{i}{\hbar}\int_{t_0}^t 2
\Ga
J_2 dt'
\ri) \ri] |\psi(t)\ran_{H_\I} ~,
\eea
where ${\hat{T}}$ denotes time-ordering.
We remark that for dimensional reasons, in Eq.(\ref{eqt0}) we need a
constant $\hbar$, with dimension of an action which from purely classical
considerations cannot be fixed in magnitude. Quantum Mechanics tells us
how to fix it. Exactly the same situation occurred in the
classical statistical mechanics of Gibbs: the precise value to be chosen
for the action quantum $2 \pi \hbar$ was evident only after quantum
theory.

The states $|\psi(t)\ran_{H}$  and $|\psi(t)\ran_{H_\I}$ satisfy the
equations:
\bea \lab{S1}
i \hbar \frac{d}{dt} |\psi(t)\ran_{H} &=& H \,|\psi(t)\ran_{H}~,
\\ \lab{S2}
i \hbar \frac{d}{dt} |\psi(t)\ran_{H_\I} &= &2 \Om
{\cal C} |\psi(t)\ran_{H_\I} \, .
\eea
Eq.(\ref{S2}) describes the 2D ``isotropic''
(or ``radial'') harmonic oscillator.
$ H_\I = 2 \Om{\cal C} $ has the  spectrum
${\cal H}^n_\I= \hbar \Om n$, $n = 0,
\pm 1, \pm 2, ...$. According to our choice for ${\cal C}$ to
be positive, only positive values of $n$ will be considered.

We can conveniently choose $t_0=t$ if we consider the integration along
the closed time path, say $C_t$, which goes from $t+i\ep$ to some
arbitrary final time $t_f$ and back to $t- i\ep$ \cite{schwinger}.
 Since $J_2$ is time
independent, we can drop the time ordering and write:
\begin{equation}
|\psi(t)\ran_{H}=
\exp\lf( i \int_{C_t}
A(t') dt' \ri)  |\psi(t)\ran_{H_\I} \, ,
\label{eq0T}
\end{equation}
where $A(t) \equiv \frac{\Ga m}{ \hbar}({\dot x}_1 x_2 - {\dot x}_2 x_1)$
and we used
the fact that $\int_{C_t}r^2= \int_{t_0}^t r^2 + \int_{t}^{t_0} r^2 =
\int_{t_0}^t r^2 - \int_{t_0}^t r^2 =0$. Note that $({\dot x}_1 x_2 -
{\dot x}_2 x_1) dt$ is the area
element in the $(x_1,x_2)$ plane enclosed by the trajectories (see Fig.1).
Consider the above expression for $t=\tau$ and $t=0$:
\bea
|\psi(\tau)\ran_{H} &=&
\exp\lf( i \int_{C_{\tau}}
A(t') dt' \ri)  |\psi(\tau)\ran_{H_\I} \, ,
 \\
|\psi(0)\ran_{H} &=&\exp\lf( i \int_{C_0}
A(t') dt' \ri)  |\psi(0)\ran_{H_\I}\, ,
\label{eq0Tbis}
\eea
where $C_{\tau}$ and $C_0$ are the time contours going from $\tau$ (or
from $0$)
to $t_f$ and back along the real line. We focus now on eigenstates of
$H$ and $H_I$.
We then get
\bea
&&\,_{H}\lan \psi(\tau) | \psi(0) \ran_{H}
\nonumber \\
&&=\,_{H_\I}\!\lan \psi(0)| \exp\lf( i \int_{C_{0\tau}}
A(t') dt' \ri)  |\psi(0)\ran_{H_\I} \nonumber \\
&& \equiv e^{i \phi} \, ,
\label{berryphase}
\eea
where the contour $C_{0\tau}$ is the one going from $t'=0$ to $t'=\tau$
and back.
Notice that the dependence on the (arbitrary) final time $t_f$ has
disappeared. One may
observe that the evolution (or dynamical) part of
the phase does not enter in $\phi$, as the integral in
Eq.(\ref{berryphase}) picks up  a purely
geometric contribution (Berry--Anandan--like phase\cite{Berry}).
The integral in (\ref{berryphase}) can be calculated
by rewriting it as a contour integral in the complex plane.
If $x_1$ and $x_2$ are analytically continued into imaginary $u$,
we may define $z= x_1 +  x_2$ and ${\bar z} = x_1 - x_2$. The hyperbola
is thus mapped onto a circle in the Gaussian plane.
We then obtain
\bea
\int_{C_{0\tau}}  A(t') dt' = -  \frac{\Ga m}{\hbar} \,
\mbox{Im} \lf[ \int_\De dz {\bar z} \ri] \, = \, \pi R^2 \frac{\ga
}{\hbar}  ~,
\eea
where $\De$ is the clockwise contour in the complex plane
with radius $R$ ($z{\bar z}=R^2$):
this is equal to the hyperbolic radius for the
envelope of the trajectories in the plane $(x_1,x_2)$.
The value of $R$ is obtained by extremizing $r(t)$. We get
\bea \lab{radius}
R = \sqrt{\frac{2 {\cal E}}{m}} \frac{1}{\Om}\, ,
\eea
where ${\cal E}$ is the initial  energy given by ${\cal
E}=\frac{1}{2} m v_0^2 + \frac{1}{2} m \Om^2 r_0^2$
(we use $r(t)= r_0 \cos \Om t +
\frac{v_0}{\Om} \sin\Om t$). Note that ${\cal E}= 0$ is not
allowed since it corresponds to the system confined to the origin, which
we exclude.
In conclusion, we get $\phi = \alpha \pi$, with dimensionless
$\alpha\equiv R^2 \frac{\ga}{\hbar}$.

\begin{figure}[t]
\centerline{\hspace{0.6cm}\epsfysize=1.85truein\epsfbox{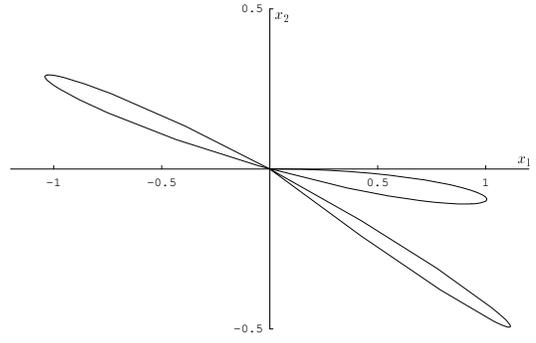}}
\vspace{0.5cm}
\caption{Trajectories for $r_0=0$ and $v_0=\Om$,
after three half-periods for $\ka =20$, $\ga=1.2$ and
$m=5$. The ratio $\int_0^{\tau/2}({\dot x}_1 x_2 - {\dot x}_2 x_1)dt
/{\cal E} =\pi\frac{\Ga}{m\Om^3}$  is preserved.}

\vspace{0.3cm} \hrule
\end{figure}

Because the physical states $|\psi\rangle$ are periodic ones, let us
focus our attention on those. Following \cite{Berry},
one may generally write
\begin{eqnarray}
|\psi(\tau)\ran &= &e^{ i\phi
- \frac{i}{\hbar}\int_{0}^{\tau}\langle \psi(t)| H |\psi(t) \rangle dt}
|\psi(0)\ran
\nonumber \\
&=&  e^{- i2\pi n} | \psi(0)\ran \, ,
\label{eqtH}
\end{eqnarray}
i.e. $\frac{ \langle \psi(\tau)| H |\psi(\tau) \rangle  }{\hbar} \tau - \phi
= 2\pi n$, $n = 0,  1,  2, \ldots $,  which by using
$\tau = \frac{2 \pi}{\Om}$ and $\phi = \alpha \pi$,
gives
\bea\lab{spectrum}
{\cal H}_{\I,e\!f\!f}^n \equiv
\langle \psi_{n}(\tau)| H |\psi_{n}(\tau) \rangle= \hbar
\Om \lf( n + \frac{\alpha}{2} \ri) ~,
\eea
where the index $n$ has been introduced to exhibit the $n$ dependence
of the state  and  the corresponding energy. ${\cal H}_{\I,e\!f\!f}^n$
gives the
effective $n$th energy level of the physical system,
namely the energy given by ${\cal H}_{\I}^n$
corrected by its interaction with the
environment. We thus see that the dissipation term $J_2$ of the
Hamiltonian  is actually
responsible for the  ``zero point energy" ($n = 0$):
$E_{0} =\frac{\hbar}{2} \Om \alpha$.

As well known, the zero point energy is the
``signature" of quantization since in Quantum Mechanics it is formally
due to the non-zero commutator of the canonically conjugate $q$ and $p$
operators. Thus dissipation manifests itself
as ``quantization". In other
words, $E_0$, which appears as the ``quantum contribution" to the spectrum
of the conservative evolution of
physical states, signals the underlying dissipative dynamics.  If we
really want to match the Quantum Mechanics zero point energy, we have to
resort to experiment, which fixes $\alpha = 1$, with $\hbar$ being the
Planck constant. In turn this fully exhibits the geometric nature
of the phase $\phi$ and gives $R_0=\frac{\hbar}{\ga}$. From
Eq.(\ref{radius}) we then see that ${\cal E}_0=\frac{1}{2}m \Om^2 R_0^2$
is the energy corresponding to $E_0$. We thus put ${\cal E}_0=E_0$, which
gives $\Om = \frac{\ga}{m}$, i.e. $\ka = 5
\frac{\ga^2}{4 m}$, consistent with the reality condition for
$\Om$.

Notice that the only free parameter of the theory is the ratio
$\frac{\ka}{m}$. We also remark that in ref. \cite{Blasone:1998xt} the phase
integral in Eq.(\ref{berryphase}), which plays there the
r\^ole of dissipative interference phase, has proved to be in fact always
non-zero in order to have quantum mechanical interference in the electron
double slit experiment. Equivalently, in the present formalism we must
always have
$x_1 \neq x_2$ (i.e. $r \neq 0$) in order to have $A(t)\neq 0 $
(cf. Eq.(\ref{eq0T})) and thus a non-zero geometric phase.

In order to better understand the dynamical r\^ole of $J_2$ we rewrite
Eq.(\ref{eqt0}) as follows
\bea\lab{eqt0U} |\psi(t)\ran_{H} =
{\hat T}\lf[ \exp\lf(i \frac{1}{\hbar}
\int_{u(t_0)}^{u(t)} 2 J_2 du'\ri) \ri]
|\psi(t)\ran_{H_\I} \, , \eea
by using $u(t) = - \Ga t$. Accordingly, we have
\bea \lab{Su}
-i \hbar \frac{\pa}{\pa u} |\psi(t)\ran_{H} = 2J_{2}
|\psi(t)\ran_{H} \, .
\eea
We thus see that $2 J_2$ is responsible for shifts (translations) in the
$u$ variable, as is to be expected since $2 J_{2} = p_{u}$ (cf.
Eq.(\ref{j2a})). In operatorial notation we can write indeed $p_{u} =- i
\hbar \frac{\pa}{\pa u}$. Then, in full generality,
Eq.(\ref{thermalcondition}) defines families of physical states,
representing stable, periodic trajectories (cf. Eq.(\ref{17})).
$2 J_{2}$ implements transition from family to family,
according to Eq.(\ref{Su}). Eq.(\ref{S1}) can be then rewritten as
\bea \lab{S11}
i \hbar \frac{d}{dt} |\psi(t)\ran_{H} = i \hbar \frac{\pa}{\pa t}
|\psi(t)\ran_{H} + i \hbar \frac{du}{dt}\frac{\pa}{\pa u}
|\psi(t)\ran_{H}\, ,
\eea
where the first term on the r.h.s. denotes of course derivative with
respect to the explicit time dependence of the state. The dissipation
contribution to the energy is thus described by the ``translations" in the
$u$ variable.  It is then interesting to consider the derivative
\bea\lab{T}
\frac{\pa S}{\pa U} = \frac{1}{T}\,.
\eea
>From Eq.(\ref{pqham}), by using $S \equiv \frac{2 J_{2}}{\hbar}$ and
$U \equiv  2 \Om {\cal C}$, we obtain $T = \hbar \Ga$.
Eq. (\ref{T}) is the defining relation for temperature in thermodynamics
\cite{Lan} (with $k_B = 1$) so that one could formally regard $\hbar
\Ga$ (which
dimensionally is an energy) as the temperature, provided the
dimensionless quantity $S$ is identified with the entropy.
In such a case, the ``full Hamiltonian'' Eq.(\ref{pqham})
plays the r{\^o}le of the free energy
${\cal F}$: $H = 2 \Om {\cal C}
- (\hbar \Ga) \frac{2 J_2}{\hbar} = U - TS = {\cal
F}$.  Thus $2 \Ga J_{2}$
represents the heat contribution in $H$ (or $\cal F$). Of course,
consistently, $\lf. \frac{\pa {\cal F} }{\pa T }\ri|_\Om = - \frac{2
J_2}{\hbar}$.  In conclusion $\frac{2 J_{2}}{\hbar}$ behaves as
the entropy, which is not surprising since it controls the dissipative
(thus irreversible) part of the dynamics.

We can also take the derivative of ${\cal F}$ (keeping $T$ fixed) with
respect to $\Om$. We then have
\bea
\lf. \frac{\pa {\cal F}}{\pa \Om}\ri|_T  =
\lf. \frac{\pa U}{\pa \Om}\ri|_T  =m r^2 \Om\, ,
\eea
which is the angular momentum: this is to be expected since it is the
conjugate variable of the angular velocity $\Om$. It is also
suggestive that the temperature $\hbar \Ga$ is actually given by
the background zero point energy: $\hbar \Ga = \frac{\hbar \Om}{2}$.

In the light of the above results,
the condition (\ref{thermalcondition}) can be then
interpreted
as a condition for  an adiabatic physical system. $\frac{2
J_{2}}{\hbar}$ might
be viewed as an analogue of the Kolmogorov--Sinai entropy for chaotic
dynamical systems.

Finally, we remark that the thermodynamical picture above outlined is
also consistent with the results on the canonical quantization of
dissipative system in quantum field theory (QFT) presented in
ref. \cite{Celeghini:1992yv}.
The transitions between unitarily inequivalent representations of
the canonical commutation relations induced by the entropy operator in
QFT correspond to the transitions
between families of stable trajectories induced
by the entropy $\frac{2 J_{2}}{\hbar}$ in the present paper.

\vspace{0.5cm}

G.V. is very grateful to Alan Widom for long,
illuminating discussions and P.J. wishes to thank John C. Taylor for many
useful comments.


\end{document}